\documentclass[10pt]{article}
\usepackage[dvips]{graphicx}

\usepackage{epsfig}
\usepackage{graphicx}
\usepackage{indentfirst}
\usepackage{amsmath}
\usepackage{amsfonts}
\usepackage{amssymb}

\begin{document}

\begin{center}
{\Large\bf  Particle creation in flat FRW Universe in the framework of $f(T)$ gravity\\}
\medskip

  M. R. Setare\footnote{e-mail:
rezakord@ipm.ir}\\
{ \it Department of Campus of Bijar, University of  Kurdistan  \\
Bijar, IRAN.}
\\and\\
 M. J. S. Houndjo\footnote{e-mail:
sthoundjo@yahoo.fr}\\
 { \it Departamento de Engenharia e Ci\^{e}ncias Naturais- CEUNES\\
Universidade Federal do Esp\'irito Santo\\
CEP 29933-415 - S\~ao Mateus-ES, Brazil}\\
{\it 
Institut de Math\'ematiques et de Sciences Physiques (IMSP)\\
01 BP 613 Porto-Novo, B\'enin}\\
\medskip
\end{center}

\begin{abstract}
We study particle production aspect in the flat FRW universe in the framework of $f(T)$ gravity. The matter is minimally coupled with the gravity  and an exact power-law solution is obtained by solving the Friedmann equations, and matter is assumed to be minimally coupled with gravitation.   The torsion scalar $T$ appears to plays the same role as the the curvature (Ricci scalar) in the the General Relativity (GR) and its modified  theories $f(R)$. Specially in phantom phase, we observe that the vacuum state corresponds to a vanishing torsion scalar and  particle production becomes
 important   as the torsion scalar diverges. This  aspect, not only provides the equivalence between the teleparallel gravity and the GR, but also between their respective modified versions, f(T) and f(R), in the view of massless particle production phenomenon when the matter is minimally coupled with the gravity.  However, when the gravitational and scalar fields are not minimally coupled, it appears that this similarity  between the teleparallel and GR may break down, due to the fact that the torsion scalar has no longer the same time dependent expression as the Ricci scalar.
\end{abstract}

Pacs numbers: 98.80.k, 98.80.Es

\section{Introduction}
The possibility of particle production due to space-time curvature has been discussed by Schrodinger \cite{1}, while other early work is due to DeWitt \cite{2}, and Imamura \cite{3}. The first thorough treatment of particle production by an external gravitational filed was given by Parker \cite{4,5}. In flat space-time, Lorentz invariance is a guide which generally allows to identify a unique vacuum state for the theory. However, in curved space-time, we do not have Lorentz symmetry. In general,
there does not exist a unique vacuum state in a curved spacetime. As a result, the concept of particles becomes ambiguous, and the problem of the physical interpretation becomes much more difficult \cite{6,7}. The creation of particles from the vacuum
takes place due to the interaction with dynamical external constraints. For example the motion of a single reflecting boundary (mirror) can create particles \cite{6}, the creation of particles by time-dependent external gravitational field is another example of dynamical external constraints. \par
Nowadays it is strongly believed that the universe is experiencing an accelerated expansion. Recent observations from type Ia supernovae \cite{SN} in associated with Large Scale Structure \cite{LSS} and Cosmic Microwave Background anisotropies \cite{CMB} have provided main evidence for this cosmic acceleration. It seems that some unknown energy components ( dark energy) with negative pressure are responsible for this late-time acceleration \cite{sah}. However, understanding the nature of dark energy is one of the fundamental problems of modern theoretical cosmology. An alternative approach to accommodate dark energy is modifying the general theory of relativity on large scales. Among these theories, scalar-tensor theories \cite{8}, f(R) gravity \cite{9}, DGP braneworld gravity \cite{10} and string-inspired theories
\cite{11} are studied extensively. On the other hand, a theory of $f(T)$ gravity has recently been received attention. In \cite{daoudah1,daoudah2}, new spherically symmetric solutions of black hole and wormhole are obtained. The reconstruction of $f(T)$  gravity has been performed  according to the holographic dark energy \cite{daoudah4}. Models based on modified teleparallel gravity were presented, in one hand, as an alternative to inflationary models \cite{16, 17}, and on the
other hand, as an alternative to dark energy models \cite{18}.\par

In this paper we consider the particle creation in the spatially flat Robertson-Walker space-time in the framework of $f(T)$ gravity. The function $f(T)$ is written as a sum of the teleparallel term $T$, and an arbitrary function $g(T)$, i.e., $f(T)=T+g(T)$. We obtain the Bogoliubov coefficients, after that the number of particles produced and the energy related to those can be explicitly found. We observe that the torsion scalar plays the same role in $f(T)$ gravity as in GR and $f(R)$ theories, in the view of particle production, at least when the scalar field is minimally coupled with gravity. The appearance of future finite-time singularities is also checked and the quantum effects due to particle creation analysed at singularity time. We perceive that in the minimally coupling case,  quantum effects are the same in  $f(T)$ gravity  as in GR and $f(R)$ gravity, and then, quantum effects may not avoid the occurrence of the big rip. However,  we observe that when gravitational and scalar fields are not minimally coupled, the   similarity between teleparallel and GR may break down, in the view of particle production, despite the torsion scalar remains the same, since the time dependent expression of the torsion scalar $T$ is not the same as that of the Ricci scalar $R$ in GR.\par
The paper is organized as follows. In Sec. $2$,  the field equations are presented in the FRW universe. In Sec. $3$, particle production phenomenon in an expanding universe is extensively discussed. Finally, in Sec. 4, we present the conclusion.

\section{Field equations within FRW metric.}
Let us consider the action  for the theory of modified gravity based on a
modification of the teleparallel equivalent of GR, namely $f(T)$ theory of gravity, coupled with matter $\mathcal{L}_M$, given by \cite{18}-\cite{20}
\begin{equation}\label{1}
S=\frac{1}{16\pi G}\int d^4x\, e \left[T+g(T)+\mathcal{L}_M\right]_,
\end{equation}
where $e=det(e^i_{\mu})=\sqrt{-g}$, and $g(T)$ denotes an
arbitrary function of $T$. In other words, the function $f(T)$ is assumed to be a sum of the teleparallel gravity term $T$ and an arbitrary function $g(T)$, i.e., $f(T)=T+g(T)$.\par

The torsion $T$ is defined as follows
\begin{equation}\label{2}
T=S^{\:\:\:\mu \nu}_{\lambda} T_{\:\:\:\mu \nu}^{\lambda},
\end{equation}
where
$$
T_{\:\:\:\mu \nu}^{\lambda}=e_i^{\lambda}(\partial_{\mu}
e^i_{\nu}-\partial_{\nu} e^i_{\mu}),
$$
$$
S^{\:\:\:\mu \nu}_{\lambda}=\frac{1}{2}(K^{\mu
\nu}_{\:\:\:\:\:\lambda}+\delta^{\mu}_{\lambda} T^{\theta
\nu}_{\:\:\:\theta}-\delta^{\nu}_{\lambda} T^{\theta
\mu}_{\:\:\:\theta}),
$$
and the contorsion $K^{\mu \nu}_{\:\:\:\:\:\lambda}$ is given by
$$
K^{\mu \nu}_{\:\:\:\:\:\lambda}=-\frac{1}{2}(T^{\mu
\nu}_{\:\:\:\:\:\lambda}-T^{\nu \mu}_{\:\:\:\:\:\lambda}-T^{\:\:\:\mu
\nu}_{\lambda}).
$$
The field equations are obtained by varying the action with
respect to vierbein $e^i_{\mu}$ and one gets
\begin {equation}\label{3}
e^{-1}\partial_{\mu}(e S^{\:\:\:\mu
\nu}_{i})(1+g_T)-e_i^{\:\sigma}T_{\:\:\:\mu
\sigma}^{\lambda}S^{\:\:\:\nu \mu}_{\lambda}g_T +S^{\:\:\:\mu
\nu}_{i}\partial_{\mu}(T)g_{TT}-\frac{1}{4}e_{\:i}^{\nu}
(1+g(T))=4 \pi G e_i^{\:\lambda}\mathcal{T}_{\lambda}^{\:\:\nu},
\end{equation}
where $g_T$ and $g_{TT}$ are the first and second derivatives of $g(T)$ with respect to $T$. Here, $\mathcal{T}_{\lambda\nu}$ denotes the stress tensor and may not be confused with the torsion.  Now, we take the usual spatially-flat metric of Friedmann-Robertson-Walker (FRW) universe, in agreement with observations
\begin {equation}\label{4}
ds^{2}=dt^{2}-a(t)^{2}\sum^{3}_{i=1}(dx^{i})^{2},
\end{equation}
where $a(t)$ is the scale factor as function of the cosmic time $t$. Moreover, we assume the background to be a perfect fluid. Using the Friedmann-Robertson-Walker metric and the perfect fluid matter in the teleparallel Lagrangian (\ref{2}) and the field equations (\ref{3}), one obtains
\begin {equation}\label{4'}
T=-6H^2,
\end{equation}
\begin {equation}\label{5}
H^2=\frac{8 \pi G \rho}{3}-\frac{1}{6}g-2H^2g_T,
\end{equation}
\begin {equation}\label{5'}
\dot{H}=-\frac{4 \pi G (\rho+p)}{1+g_T-12H^2g_{TT}},
\end{equation}
where $\rho$ and $p$ denote the matter density and pressure respectively, and the Hubble parameter $H$ is defined by
$H=\dot{a}/a$.

In the FRW universe, the energy conservation law can be expressed as the standard continuity equation
\begin {equation}\label{6}
\dot{\rho}+3H(\rho+p)=\dot{\rho}+3H(1+w)\rho=0\ ,
\end{equation}
where we used the barotropic equation of state $p=w\rho$  relating  the ordinary pressure with the  ordinary energy density. The same continuity equation should be written for the total content of the universe as well as the dark fluid.  Let us define the conformal time $\eta$ from the cosmic one, such that $dt=a(\eta)d\eta$. Thus (\ref{5}) and (\ref{5'}) can be written as
\begin{eqnarray}
\frac{a'^2}{a^4}&=&\frac{8\pi G}{3}\rho_{eff}\,\,\,,\label{5conf}\\
\frac{a''}{a^3}-2\frac{a'^2}{a^4}&=&-\frac{8\pi G}{2}\left(\rho_{eff}+p_{eff}\right)\,\,,\label{5'conf}
\end{eqnarray}
where $\rho_{eff}$ and $p_{eff}$ are the effective energy density and effective pressure, respectively, and the right hand sides of (\ref{5}) and (\ref{5conf}) are equal, respectively for (\ref{5'}) and (\ref{5'conf}). By combining (\ref{5conf}) and (\ref{5'conf}), one gets the power-law solution for the scale factor in terms of  $\eta$ as
\begin{equation}\label{9}
a(\eta)=a_0\left|\eta\right|^{b}\,,\quad b=\frac{2}{1+3\omega_{eff}}\,\,\,,
\end{equation}
where the parameter $\omega_{eff}=p_{eff}/\rho_{eff}$ depends on the algebraic function $g(T)$ and its first and second derivatives, as well as the ordinary components $\rho$ and $p$. Note that if $\omega_{eff}>-1/3$ the universe evolves so that $0<\eta<\infty$ ($0<t<\infty$) corresponding to a decelerating universe. For $\omega_{eff}<-1/3$, the interval is $-\infty<\eta<0_-$, meaning that $0<t<\infty$ when $-1/3>\omega_{eff}>-1$, and this corresponds to an accelerating universe but with the quintessence feature. But for $\omega_{eff}<-1$, the interval of $\eta$ corresponds to $-\infty<t<0_-$ for the cosmic one, and the universe is dominated by a phantom fluid: the scale factor tends to $\infty$ when $\eta\rightarrow 0_-$ ($t\rightarrow 0_-$). \par
By using the continuity equation for the total content of the universe, the effective energy density can be written as

\begin {equation}\label{7}
\rho_{eff}(t)=\bar{\rho}\left|\eta\right|^{-6(1+\omega_{eff})/(1+3\omega_{eff})}\,\,\,,
\end{equation}
where $\bar{\rho}$ is a positive constant.
Also, using (\ref{9}), Eq.(\ref{4'}) leads  to the following result
\begin {equation}
T=-6\frac{b^2}{a_0^2}\left|\eta\right|^{-6(1+\omega_{eff})/(1+3\omega_{eff})}               <0\quad. \label{torsionexpress}
\end{equation}
We see from (\ref{torsionexpress}) that if $\omega_{eff}>-1/3$, at early time $\eta\rightarrow 0$, the exponent of $|\eta|$ in (\ref{torsionexpress}) is negative and the torsion scalar diverges, $T\rightarrow -\infty$, while for late-time $\eta\rightarrow \infty$, the torsion scalar goes to zero ($T\rightarrow 0_-$). For 
$-1<\omega_{eff}<-1/3$, the exponent of $|\eta|$ is positive and at the early time $\eta\rightarrow -\infty$, the torsion scalar also diverges ($T\rightarrow -\infty$), while at late-time $\eta\rightarrow 0_-$, the torsion scalar goes to zero. In the case where $\omega_{eff}<-1$, the exponent of $|\eta|$ is negative so that at early time $\eta\rightarrow -\infty$, the torsion scalar goes to zero ($T\rightarrow 0_-$), while at late-time $\eta\rightarrow 0_-$, the torsion scalar diverges $T\rightarrow -\infty$.
Thus, we can conclude that for an expanded decelerated universe the torsion scalar is initially strong and the conformal time evolves it goes to zero. For an expanded accelerating universe in quintessence phase,  the torsion scalar is still initially strong and as the conformal time evolves, it goes to zero. However, the opposite occurs in an expanded accelerating universe in phantom phase, i.e., the torsion scalar is initially null and as the conformal time increases it diverges as the late-time is approached.
 
\section{Particle production in an expanding universe}

Let us denote by $\phi$ a scalar field of mass $M$ and $a(\eta)$ the scale factor depending on the conformal time $\eta$ for spatially flat homogeneous and isotropic FRW spacetime. The line element in terms of the conformal time is written as follows
\begin{eqnarray}\label{vin1}
ds^2=a^2(\eta)\left( d\eta^2-dx^2-dy^2-dz^2\right)\,\,.
\end{eqnarray}
The Lagrangian that describes the minimal coupling between the massive scalar field with the gravitational in conformal spacetime is
\begin{eqnarray}\label{vin2}
\mathcal{L}_M=\frac{1}{2}a^2\eta^{\mu\nu}(\partial_\mu\phi)(\partial_\nu\phi)-\frac{1}{2}a^4M^2\phi^2\quad,
\end{eqnarray}
where $\eta_{\mu\nu}$ is the metric in  Minkowski space. The corresponding field equations can be written as
\begin{eqnarray}\label{vin3}
\frac{1}{a^2}\eta^{\mu\nu}\partial_{\mu}(a^2\partial_\nu)+a^2M^2\phi=0\quad.
\end{eqnarray}
The real scalar field can be decomposed into the modes as
\begin{eqnarray}\label{vin4}
\phi(\eta,\vec{x})=\frac{1}{a}\int \frac{d^3k}{(2\pi)^{3/2}} e^{-i\vec{k}\vec{x}}\left[\chi_k(\eta)a_{\vec k}+\chi^{\ast}_{k}(\eta)a^{\dagger}_{-\vec k}\right]\quad.
\end{eqnarray}
One can observe that the modes functions satisfy the Klein-Gordon equation
\begin{eqnarray}\label{vin5}
\chi^{\prime\prime}_k+\left(k^2+M^2a^2-\frac{a^{\prime\prime}}{a}\right)\chi_k=0\quad,
\end{eqnarray}
where we used $\phi=a\chi$, and the prime denotes the derivative with respect to the conformal time $\eta$.
The modes $\chi_k$ satisfy to the Wronskian relation
\begin{eqnarray}\label{vin6}
\chi_k(\eta)\chi^{\ast\,\prime}_{k}(\eta) - \chi^{\ast}(\eta)\chi^{\prime}_{k}(\eta)=-i\quad.
\end{eqnarray}
The system can be quantized in a standard fashion by using it as an operator, imposing the equal-time commutation relations
\begin{eqnarray}\label{vin7}
\left[\chi(\eta,\vec{x}),\pi(\eta,\vec{x}^{\,\prime})\right]= i\delta^3(\vec{x}-\vec{x}^{\,\prime})\,\,,
\end{eqnarray}
where $\pi=d\chi/d\eta\equiv \chi^{\,\prime}$ is the canonical momentum. Therefore, the operators $a_{\vec{k}}$ and $a_{\vec{k}}^{\dagger}$ satisfy the usual commutation relations
\begin{eqnarray}\label{vin8}
[a_{\vec{k}},a^{\dagger}_{\vec{k}^{\prime}}]=\delta^3(\vec{k}-\vec{k}^{\,\prime})\,\quad,\quad[a_{\vec{k}},a_{\vec{k}^{\prime}}]=[a^{\dagger}_{\vec{k}},a^{\dagger}_{\vec{k}^{\prime}}]=0 \,\,.
\end{eqnarray}
The vacuum state can be defined as the state for which the following condition is satisfied
\begin{eqnarray}\label{vin9}
a_{\vec{k}}\left|0\right\rangle=0\quad\forall\,\, k\quad.
\end{eqnarray}
One may also construct other states from the vacuum, by acting on it various combinations of creation operators $a^{\dagger}_{\vec{k}}$.\par
The Hamiltonian that corresponds to  (\ref{vin2}) is
\begin{eqnarray}\label{vin10}
H=\frac{1}{2}\int d^3x\left\{a^2\left[(\phi^{\prime})^2+(\nabla \phi)^2\right]+a^4M^2\phi^2\right\}\quad.
\end{eqnarray}
By using (\ref{vin4}), we re-write the Hamiltonian  in terms of the creation and annihilation operators and the mode functions as
\begin{eqnarray}\label{vin11}
H=\frac{1}{2}\int d^3k \left[ P_k(\eta)a_{\vec k}a_{-\vec k}+P^{\ast}_{k}(\eta)a^{\dagger}_{\vec k}a^{\dagger}_{-\vec k}+Q_{k}(\eta)\left(a_{\vec k}a^{\dagger}_{\vec k}+a^{\dagger}_{\vec k}a_{\vec k}\right)\right]\quad,
\end{eqnarray}
where $P_k(\eta)$ and $Q_{k}(\eta)$ are defined by
\begin{eqnarray}
P_k&=&\left(-\frac{a^{\prime}}{a}\chi_k+\chi^{\prime}_k\right)^2+\omega_k^2\chi_k^2\quad, \label{vin12}\\
Q_k&=&\left(-\frac{a^{\prime}}{a}\chi_k+\chi_{k}^{\prime}\right)\left(-\frac{a^\prime}{a}\chi^{\ast}_{k}+\chi^{\ast\,\prime}_{k}\right)+\omega^{2}_{k}\chi_k\chi^{\ast}_{k}
\quad,\label{vin13}
\end{eqnarray}
with $\omega_k^2=k^2+M^2a^2 $.\par
In general,  different complete sets of modes can be used for constructing the field and each of them possesses its own vacuum state. If we label one of these set of modes by  $\bar{\chi}(\eta)$, then, the field can be written in terms of these modes as
\begin{eqnarray}\label{vin14}
\phi(\eta,\vec{x})=\frac{1}{a}\int \frac{d^3k}{(2\pi)^{3/2}} e^{-i\vec{k}\vec{x}}\left[\bar{\chi}_k(\eta)\bar{a}_{\vec k}+\bar{\chi}^{\ast}_{k}(\eta)\bar{a}^{\dagger}_{-\vec k}\right]\quad,
\end{eqnarray}
and the corresponding vacuum state is such that $\bar{a}_{\vec{k}}\left|0\right\rangle=0$ for all $\vec{k}$. The Hamiltonian in this case is written as
\begin{eqnarray}\label{vin15}
H=\frac{1}{2}\int d^3k \left[ \bar{P}_k(\eta)\bar{a}_{\vec k}\bar{a}_{-\vec k}+\bar{P}^{\ast}_{k}(\eta)\bar{a}^{\dagger}_{\vec k}\bar{a}^{\dagger}_{-\vec k}+\bar{Q}_{k}(\eta)\left(\bar{a}_{\vec k}\bar{a}^{\dagger}_{\vec k}+\bar{a}^{\dagger}_{\vec k}\bar{a}_{\vec k}\right)\right]\quad,
\end{eqnarray}
where $\bar{P}_k$ and $\bar{Q}_k$ are similar expressions as in
(\ref{vin12}) and (\ref{vin13}) respectively, replacing
$\chi_k$ by $\bar{\chi}_k$. Taken in consideration the  completeness, the two sets
of modes $\chi_{k}$ and $\bar{\chi}_{k}$ are related by the
Bogoliubov transformations that diagonalize the Hamiltonian i.e.,
$\bar{P}_k(\eta)=0$, satisfying the  relation
\begin{eqnarray}\label{vin16}
\bar{\chi}_{k}(\eta)=\gamma_k\chi_{k}(\eta)+\beta_k\chi^{\ast}_{k}(\eta)\,\,,
\end{eqnarray}
with the  normalization condition
$|\gamma_k(\eta)|^2-|\beta_k(\eta)|^2=1$, where $\gamma_k$ and
$\beta_k$ are called Bogoliubov coefficients. We can now  compare the two vacuum states by noting that the number
operator for the barred states is $\bar{N}\equiv \int
d^3k\bar{a}^{\dagger}_{\vec{k}}\bar{a}_{\vec{k}}$. Computing  its expectation value with respect to the unbarred vacuum, we get
\begin{eqnarray}\label{vin17}
\left\langle 0\right|\bar{N}\left|0\right\rangle = \int d^3k\left|\beta_k\right|^2\,\,.
\end{eqnarray}
Hence, the number of barred particles in the unbarred vacuum in the mode $\vec{k}$ is $\left|\beta_k\right|^2$. In the same way, the number of unbarred particles in the barred vacuum in the mode $\vec{k}$ is $\left|\beta_k\right|^2$. Since the diagonalization imposes $\bar{P}_k(\eta)=0$, using (\ref{vin12}) with $\chi_k$ substituted  by $\bar{\chi}_k$ , one gets
\begin{eqnarray}\label{vin18}
-\frac{a^{\prime}}{a}\bar{\chi}_k+\bar{\chi}^{\,\prime}_k=-i\omega_k\bar{\chi}_k\quad.
\end{eqnarray}
Putting this result in the expression of $\bar{Q}_k(\eta)$, we obtain
\begin{eqnarray}\label{vin19}
\bar{Q}_k(\eta) = 2\omega_{k}^{2}(\eta)|\bar{\chi}_k|^2\quad.
\end{eqnarray}
By using (\ref{vin11}) and (\ref{vin15}), one obtains
\begin{eqnarray}\label{vin20}
\left\langle 0\right|2\bar{a}^{\dagger}_{\vec k}(\eta)\bar{a}_{\vec k}(\eta)+1\left|0\right\rangle =\frac{Q_k(\eta)}{\bar{Q}_k(\eta)}\quad,
\end{eqnarray}
and therefore
\begin{eqnarray}\label{vin21}
|\beta_k(\eta)|^2= \frac{1}{2}\frac{Q_k(\eta)}{\bar{Q}_k(\eta)}-\frac{1}{2}\quad.
\end{eqnarray}
Combining now (\ref{vin13}) and (\ref{vin19}),    (\ref{vin21}) is rewritten as
\begin{eqnarray}\label{vin22}
|\beta_k(\eta)|^2=\frac{1}{4}\frac{\left[\left(\frac{a^\prime}{a}\right)^2+\omega^2_k(\eta)\right]|\chi_k |^2-\frac{a^\prime}{a}\left(\chi_k\chi^{\ast\,\prime}_k+\chi^{\prime}_{k}\chi^{\ast}_{k}\right)+|\chi^{\prime}_{k}|^2}{\omega^2_k(\eta)|\bar{\chi}_k|^2}-\frac{1}{2}\quad.
\end{eqnarray}
There are two important aspects to be understood about the vacuum state in curved spacetime. The first is that, in general,  it is uncertain what criteria should be used for fixing the correct vacuum. Note that the criteria used in Minkowski space, as Lorentz invariance and positive frequency with respect to the timelike Killing vector, no longer apply in curved spacetime. The second aspect is that, if there is a "natural"  choice of vacuum state when the spacetime begins, in general, it does not correspond to the natural choice of vacuum when the spacetime ends. Therefore, the "in" vacuum state and the "out" vacuum state are different. This means that  the bar vacuum may contain  particles of the unbar vacuum and vice-versa. Then, particle creation occurs as can be seen from (\ref{vin22}).\par

Now, in order to calculate $|\beta_k(\eta)|^2$  we have to find
$\chi_k(\eta)$ from (\ref{vin5}) and $\bar{\chi}_k(\eta)$
from (\ref{vin18}).  
Making use of (\ref{9}) in (\ref{vin5}), the general solution can be written in terms of Hankel functions as
\begin{eqnarray}\label{vin51}
\chi_k(\eta)=\frac {\sqrt{\pi(|\eta|)}}{2}\left[A_kH^{(1)}_\nu(k|\eta|)+B_kH^{(2)}_\nu(k|\eta|)\right]\,\,, \quad \quad \nu=\sqrt{\frac{1}{4}+b(b-1)}\,\,\,,
\end{eqnarray}
where $A_k$ and $B_k$ are constants to be determined. By using  the Wronskian relation
\begin{eqnarray}\label{vin52}
zH^{(2)}_{\nu}(z)\partial_{z}H^{(1)}_{\nu}(z)-zH^{(1)}_{\nu}(z)\partial_{z}H^{(2)}_{\nu}(z)
= \frac{4i}{\pi} \quad ,
\end{eqnarray}
 and  imposing the orthonormalization of the modes, one obtains
\begin{eqnarray}\label{vin53}
\left|B_k\right|^2 - \left|A_{k}\right|^{2} = 1 \quad .
\end{eqnarray}
We fix the  initial vacuum state, using the Bunch-Davies state
\cite{article1, bunch} by the choice $A_k=0$ and $B_k=1$, and the
solution (\ref{vin51}) becomes
\begin{eqnarray}\label{vin54}
\chi_k(\eta)=\frac {\sqrt{\pi|\eta|}}{2}H^{(2)}_\nu(k|\eta|)\,\,\,.
\end{eqnarray}
 Note that as $\eta \rightarrow -\infty$ ($\omega_{eff}<-1/3$), or as $\eta \rightarrow \infty$ ($\omega_{eff}>-1/3$), the solution (\ref{vin54}) reduces to the Minkowski one in which particle production phenomenon does not occur.\par
Now we can proceed to the calculation of $|\beta_k(\eta)|^2$ through the expression (\ref{vin22}). Since we are dealing with a minimally coupled  massless scalar field,  the expression (\ref{vin22}) becomes
\begin{eqnarray}\label{vin55}
|\beta_k(\eta)|^2=\frac{1}{4}\frac{\left(\frac{b^2}{\eta^2}+k^2\right)|\chi_k |^2+\frac{b}{|\eta|}\left(\chi_k\chi^{\ast\,\prime}_k+\chi^{\prime}_{k}\chi^{\ast}_{k}\right)+|\chi^{\prime}_{k}|^2}{k^2|\bar{\chi}_k|^2}-\frac{1}{2}\,\,\,.
\end{eqnarray}
Substituting (\ref{vin54}) into (\ref{vin55}),  we get
\begin{eqnarray}\label{vin56}
|\beta_k(\eta)|^2=\frac{\pi}{8k}\Bigg\{\left[\frac{(b+1/2)^2}{|\eta|}+k^2|\eta|\right]H^{(1)}_{\nu}(k|\eta|)H^{(2)}_{\nu}(k|\eta|)\nonumber\\
+\frac{k}{2}(b+1/2)\Bigg[H^{(1)}_{\nu}(k|\eta|)H^{(2)}_{\nu-1}(k|\eta|)+H^{(2)}_{\nu}(k|\eta|)H^{(1)}_{\nu-1}(k|\eta|)\nonumber\\
-H^{(1)}_{\nu}(k|\eta|)H^{(2)}_{\nu+1}(k|\eta|)-H^{(2)}_{\nu}(k|\eta|)H^{(1)}_{\nu+1}(k|\eta|)\Bigg]\nonumber\\
+\frac{k^2|\eta|}{4}\Bigg[H^{(1)}_{\nu-1}(k|\eta|)H^{(2)}_{\nu-1}(k|\eta|)-H^{(1)}_{\nu-1}(k|\eta|)H^{(2)}_{\nu+1}(k|\eta|)\nonumber\\
-H^{(1)}_{\nu+1}(k|\eta|)H^{(2)}_{\nu-1}(k|\eta|)+H^{(1)}_{\nu+1}(k|\eta|)H^{(2)}_{\nu+1}(k|\eta|)\Bigg]-\frac{4k}{\pi}\Bigg\}\,\,.
\end{eqnarray}
 Observe that at the early time, $\eta\rightarrow -\infty$ ($\omega_{eff}<-1/3$),   using the asymptotic expressions of the Hankel functions and (\ref{vin56}), we obtain $|\beta_k(-\infty)|^2=0$. However, this is not the case when $\omega_{eff}>-1/3$, since using the asymptotic initial limit of the Hankel functions, one gets $|\beta_k(0)|^2=+\infty$. It appears  clearly through  (\ref{vin56}) that in an expanded accelerating universe, $\omega_{eff}<-1/3$, particle production is initially null, and as time evolves it becomes important at late-time. But in an expanded decelerating universe, particle production is initially strong and becomes inefficient at late-time. This seems quick reasonable because the expanded decelerating universe is the matter dominated universe and comes before the accelerating dark energy dominated one. The end of the matter dominated phase is linked with the early of the accelerating one, and one sees clearly that the matching of the two phases on the point of view  of particle production is realized, because at the end of the decelerating phase $|\eta|\rightarrow \infty$($\omega_{eff}>-1/3$), and at the beginning of the accelerating phase $|\eta|\rightarrow\infty$ ($\omega_{eff}<-1/3$), the rate of particle production vanishes. As the conformal time grows, the evolution of the total number $N(\eta)$ of created particle per unit of space volume is \cite{grib}
\begin{eqnarray}
N(\eta)&=&\int N_k(\eta)\nonumber\\
&=& \frac{1}{2\pi^2 a^3(\eta)}\int k^2dk |\beta_k(\eta)|^2 ,\label{pcnumber}
\end{eqnarray}
 and the quantum energy density associated to massless particle production is
\begin{eqnarray}
\rho_q(\eta)=\frac{1}{(2\pi)^3a^4(\eta)}\int  k^3dk |\beta_k(\eta)|^2\,\,.
\end{eqnarray}

Thus, according to the values of the effective parameter of the equation of state $\omega_{eff}$, the phenomenon of particle production can change from the early time  to the late one. We can distinguish three different cases:\par
$\bullet$  For $\omega_{eff}>-1/3$, the conformal time is such that  $0<\eta<+\infty$  and the torsion scalar $-\infty<T<0_-$, where the rate of particle production $|\beta_k(\eta)|^2$ is initially important and decreases as the time evolves. In this situation we see that in an expanded decelerating universe, the order of magnitude of the rate of particle production is directly linked with the torsion scalar, i.e., for large torsion scalar, the rate of particle production is important and for small torsion scalar , the rate of particle production becomes small.\par 

$\bullet$ For $-1<\omega_{eff}<-1/3$, one has $-\infty<\eta<0_-$ and $-\infty<T<0_-$, whereas the rate of particle production is initially null, and increases as the time grows. Note that this is a quintessence-like situation and one can conclude in this case that the magnitude of the rate of particle production is inversely linked with the 
order of magnitude of the torsion scalar, i.e., for large torsion scalar the rate of particle production vanishes, while for small torsion scalar the rate of particle production  diverges. Thus, we can conclude that in quintessence-like universe, the rate of particle production and the torsion scalar do not follow the same order of magnitude.\par 
$\bullet$ For $\omega_{eff}<-1$, we have for the conformal time $-\infty<\eta<0_-$, whereas for the torsion scalar, $0>T>-\infty$. Here, the rate of particle production is initially null and as the time grows, it increases. We see that for small magnitude of torsion scalar, the rate of particle production is also small, while for large magnitude of torsion scalar, the rate of particle production is also large. We conclude that in the phantom-like universe, the rate of particle production is directly linked with the order of magnitude of the torsion scalar.\par
In this later case, one can observe that as the final time is approached, $\eta\rightarrow 0_-$, the scale factor $a(\eta)$, the effective energy density $\rho_{eff}(\eta)$ and effective pressure $p_{eff}(\eta)$ diverge; this is the so-called Big Rip (singularity of type-I). Do to this finite time singularity, we focus our attention to this case and try to analyse the effects of particle production around the singularity. We present in Fig.1 the evolution of the quantum energy density ( energy density of particle production), as well as it quotient over the classical one (the effective energy density) $\rho_{q}/\rho_{eff}$. We also present the graph of the evolution of the torsion scalar in this phantom case. The important and interesting feature to be notified here is the evolution of $\rho_{q}/\rho_{eff}$ when the time evolves. We see that as the conformal singularity time is approached, both the classical and quantum energy densities diverge, but particularly, the quotient $\rho_{q}/\rho_{eff}$ goes to zero. This means that as the singularity is approached, quantum effect due to particle production is inefficient with respect to  the classical effective energy density of the background. Therefore, mass-less particle production cannot avoid the Big Rip.\par

This situation is the same as that in  which the gravitational part of the action is driven by the curvature and gravitational field is minimally coupled with the scalar field. In GR the curvature scalar  reads  $R=-6a''/a^3 \propto |\eta|^{-6(1+\omega)/(1+3\omega)}$. It is easy to see that this latter is proportional to the torsion scalar. Therefore, for any value of $\omega$, the evolution of the curvature scalar will be same as that of the torsion scalar, and this, for  an expanded decelerating phase $\omega>-1/3$, an expanded accelerated quintessence phase $-1<\omega<-1/3$, and the expanded accelerating phantom phase $\omega<-1$ . Particularly, in phantom phase when the curvature goes to zero or  almost null, there is no particle production, while as the curvature increases, the particle production becomes important  \cite{livreparker,6,grib,livrefulling, livremukhanov}. \par
Other important point to be put out here is that {\it particle production aspect} for massless scalar field appears to be the same for GR, $f(R)$ \cite{vincent} and $f(T)$. In $f(T)$ theory, one can show how the algebraic function $f(T)$ can influence the particle production. Remark that the scale factor depends on the effective parameter of equation of state $\omega_{eff}=p_{eff}/\rho_{eff}$, because of the dependence of $\rho_{eff}$ and $p_{eff}$ on the torsion scalar. Then, the sign of $\omega_{eff}$ will depend on the expression of the algebraic function $f(T)$. \par
As in GR and $f(R)$ gravity, one can write the Lagrangian density $\mathcal{L}_M$ in Teleparallel  (see also \cite{jamilmomeni} where this type of Lagrangian density is used and the stability of non-minimally conformally coupled scalar field has been widely developed with interesting results), as
\begin{eqnarray}
\mathcal{L}_M=\phi_{,\mu}\phi^{,\mu}-\left(M^2+\xi T\right)\phi^2\,\,.
\end{eqnarray}
Through the Euler-Lagrangian equation, and using the Fourier space, one gets the Klein-Gordon equation for each quantum mode $k$, as
\begin{eqnarray}
\phi_k''+2\frac{a'}{a}\phi_k'+\left(k^2+Ma^2+\xi a^2 T\right)\phi_k=0\,\,\,.
\end{eqnarray}
By setting $\phi=\chi/a$, one gets the following equation
\begin{eqnarray}\label{debat}
\chi^{\prime\prime}_{k}+\left(k^2+M^2a^2-\frac{a^{\prime\prime}}{a}+\xi a^2T\right)\chi_{k}=0\,\,\,.
\end{eqnarray}
Observe that for the minimally coupling, i.e., $\xi=0$, the contribution of the torsion  in this equation disappears. Thus, we see clearly that for minimally coupling constant, the aspect of particle production is the same in $f(T)$ theory as in $f(R)$ theory. It is important to note that particle production aspect is different from the quantitative aspect of particle production. What we would like to show here is the form  of Klein-Gordon equation for both $f(R)$ and $f(T)$ theories. For minimally coupling constant, and in our case where the massless field is considered, the previous equation becomes
\begin{eqnarray}
\chi^{\prime\prime}_{k}+\left(k^2-\frac{a^{\prime\prime}}{a}\right)\chi_{k}=0\,\,\,,
\end{eqnarray}
for both $f(R)$ and $f(T)$ theories of gravity. Even, the massive field is considered within minimally coupling choice, the two theories present the same result, 
\begin{eqnarray}
\chi^{\prime\prime}_{k}+\left(k^2+M^2a^2-\frac{a^{\prime\prime}}{a}\right)\chi_{k}=0\,\,.
\end{eqnarray}
As it is well known in quantum field theory \cite{livreparker,livremukhanov,grib}, particle production depends on the effective mass, which is $M_{eff}=-a''/a$ and $M_{eff}=M^2a^2-a''/a$ respectively for massless scalar field and massive scalar field minimally coupled with the gravitational one. Particularly, for the purpose of this work where massless scalar field is considered, one  sees that in phantom phase, $\omega_{eff}<-1$, the effective mass $M_{eff}=-a''/a \propto |\eta|^{-2}$. In this phantom phase, the early corresponds to $\eta\rightarrow -\infty$, for which the effective mass vanishes, leading to Klein-Gordon equation for massless field minimally coupled with the gravitational in Minkowski's spacetime, where particle production never occurs: this is the vacuum. But as the time evolves, the effective mass $M_{eff}\propto |\eta|^{-2}$ increases, and  diverges as at the late-time $\eta\rightarrow 0_-$. This result is valid for both $f(R)$(including GR) and $f(T)$
 theory. Consequently, quantum effects due to massless particle production from minimally coupled field have to be the same in $f(T)$ as in the framework of  GR and $f(R)$ gravity, where it shown that quantum effects cannot avoid the occurrence of the big rip \cite{article1,article3,pavlov,anderson,vincent}. The same would appear in the case of the sudden singularity where quantum effect is inefficient, showing  that this singularity is also robust with respect to quantum effects \cite{article2,article4,barrow2}.\par  
 
However, it is important to note that when the scalar field is not minimally coupled with the gravitational one, the particle production seems different, at least quantitatively. Note that, in both the coupling and non-coupling case, the torsion scalar remains the same, initially null and growing as the time evolves. The capital difference that appears between the  GR and the teleparallel when the fields are not minimally coupled is about the time dependent expression of the torsion scalar $T$ in teleparallel gravity and the Ricici scalar $R$ in GR. Note that in GR and $f(R)$ gravity, the  equation (\ref{debat}) is written as
\begin{eqnarray}\label{debat1}
\chi^{\prime\prime}_{k}+\left(k^2+M^2a^2-\frac{a^{\prime\prime}}{a}+\xi a^2R\right)\chi_{k}=0\,\,\,.
\end{eqnarray}
In GR and $f(R)$, the expression  of the Ricci scalar based on the conformal time is $R=-6a^{\prime\prime}/a^3$, while in teleparallel gravity and $f(T)$ theory of gravity, the expression of the torsion scalar is $T=-6H^2=-6a^{\prime\,2}/a^4$. Then, (\ref{debat}) and (\ref{debat1}) become respectively
\begin{eqnarray}
\chi_{k}^{\prime\prime}+\left[k^2+M^2a^2-6\left(\frac{1}{6}-\xi(\eta)\right)\frac{a^{\prime\prime}}{a}\right]\chi_k=0\,\,\,\,,\quad \xi(\eta)=-\frac{a^{\prime\,2}}{aa^{\prime\prime}}\xi_c\label{debat3}
\end{eqnarray}
\begin{eqnarray}
\chi_{k}^{\prime\prime}+\left[k^2+M^2a^2-6\left(\frac{1}{6}-\xi_c\right)\frac{a^{\prime\prime}}{a}\right]\chi_k=0\,\,\,\,,\label{debat4}
\end{eqnarray}
where $\xi_c$ denotes here the coupling constant commonly know in GR and $\xi(\eta)$ is the coupling constant in teleparallel gravity..  It appears clearly that if $\xi_c=0$ (the minimally coupling case), $\xi(\eta)=0$, and the equations (\ref{debat3}) and (\ref{debat4}) are equivalent. However, if the non-minimally coupling is considered, the two equations are not equivalent and consequently their solutions should not be the same. Since the solution of the equation of motion is the most necessary element for  calculating the Bogoliubov coefficients, the results must be different.  Now, looking for the equations (\ref{debat3}) and (\ref{debat4}), it can be concluded that in  $f(T)$ gravity, the coupling parameter between the scalar and gravitational  fields is a {\it running coupling constant}.
Then, in general, in the view of  particle production, the equivalence between the teleparallel and the GR no longer occurs when the scalar field is non-minimally coupled with the gravitational one.

\section{Conclusion}
 In this paper we have studied particle creation in FRW universe
in the framework of $f(T)$ gravity, considering massless scalar field minimally coupled with the gravitational one. Since the case when the spatial
sections are flat is slightly simpler,  we have restricted our attention to this  later in this work. By solving the generalized equations of Friedmann, one gets a power-law solution for the scale factor, which depends on an effective parameter of equation of state $\omega_{eff}$. In general, we see that in an expanded decelerating universe, $\omega_{eff}>-1/3$, the conformal time $\eta$ goes from $0$ to $\infty$, while the torsion scalar $T$ goes from $-\infty$ to zero. In this case, the rate of particle production $|\beta_k(\eta)|^2$ in initially strong and as the conformal time evolves it decreases and tends to zero as the late-time is approached. For $-1<\omega_{eff}<-1/3$, the quintessence-like phase, the conformal time goes from $-\infty$ to zero, and similarly, the torsion scalar goes from $-\infty$ to zero. But in this case, the rate of particle production is initially null and as the time evolves it increases and diverges as the late-time is approached. The final and interesting case is the phantom-like case where $\omega_{eff}<-1$. Here, the conformal time goes from $-\infty$ to zero whereas the torsion scalar is initially null and decreases to $-\infty$ as the time evolves. In this case, particle production is initially null, increases as the conformal time evolves and diverges at the late-time. We conclude that the rate of particle production  is directly connected to the order of magnitude of the torsion scalar in the expanded decelerating and expanded accelerating phantom universes. However, in quintessence-like universe, the situation is different: as the torsion scalar is large, the rate of particle production is inefficient and for small torsion scalar, the rate of particle production becomes important.\par
Since in the phantom-like case, the Big Rip (the singularity of type-I) may appear, we analyse the effects of particle production around the singularity. The result shows that the energy density due to particle production diverges as the singularity time is approached, but is not enough to avoid the occurrence of the singularity.\par 
Moreover, we present the comparison of particle production aspect between $f(R)$ and $f(T)$ in general. Since the interpretation of quantum particle production is based on the effective mass $M_{eff}$, we shown that particle production aspect is the same for both GR and tele-parallel theory, and this may be extended to both $f(R)$ and $f(T)$ theories, for massless scalar field minimally coupled with the gravitational one. However, when the scalar field is non-minimally coupled with the gravitational one, this equivalence breaks down for the GR and tele-parallel theory, and consequently for $f(R)$ and $f(T)$. Comparing with the GR (respectively $f(R)$ theory) where the coupling constant is effectively constant, the tele-parallel theory (respectively $f(T)$ theory) presents a running coupling constant.

\par
\vspace{1cm}
{\bf Acknowledgement}: MJSH thanks  CNPq/FAPES for financial support.


\begin{center}
\begin{figure}[t]
\begin{minipage}[t]{0.45\linewidth}
\includegraphics[width=\linewidth]{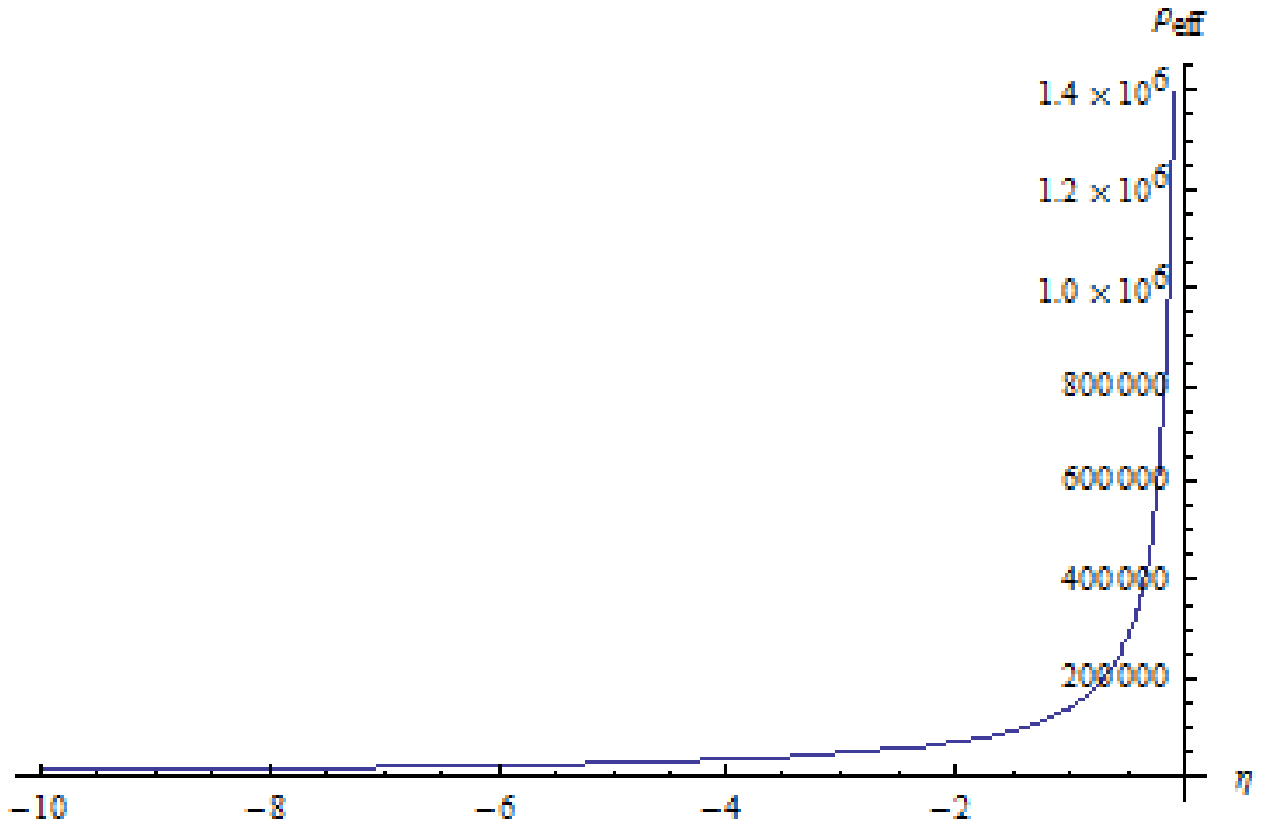}
\end{minipage} \hfill
\begin{minipage}[t]{0.45\linewidth}
\includegraphics[width=\linewidth]{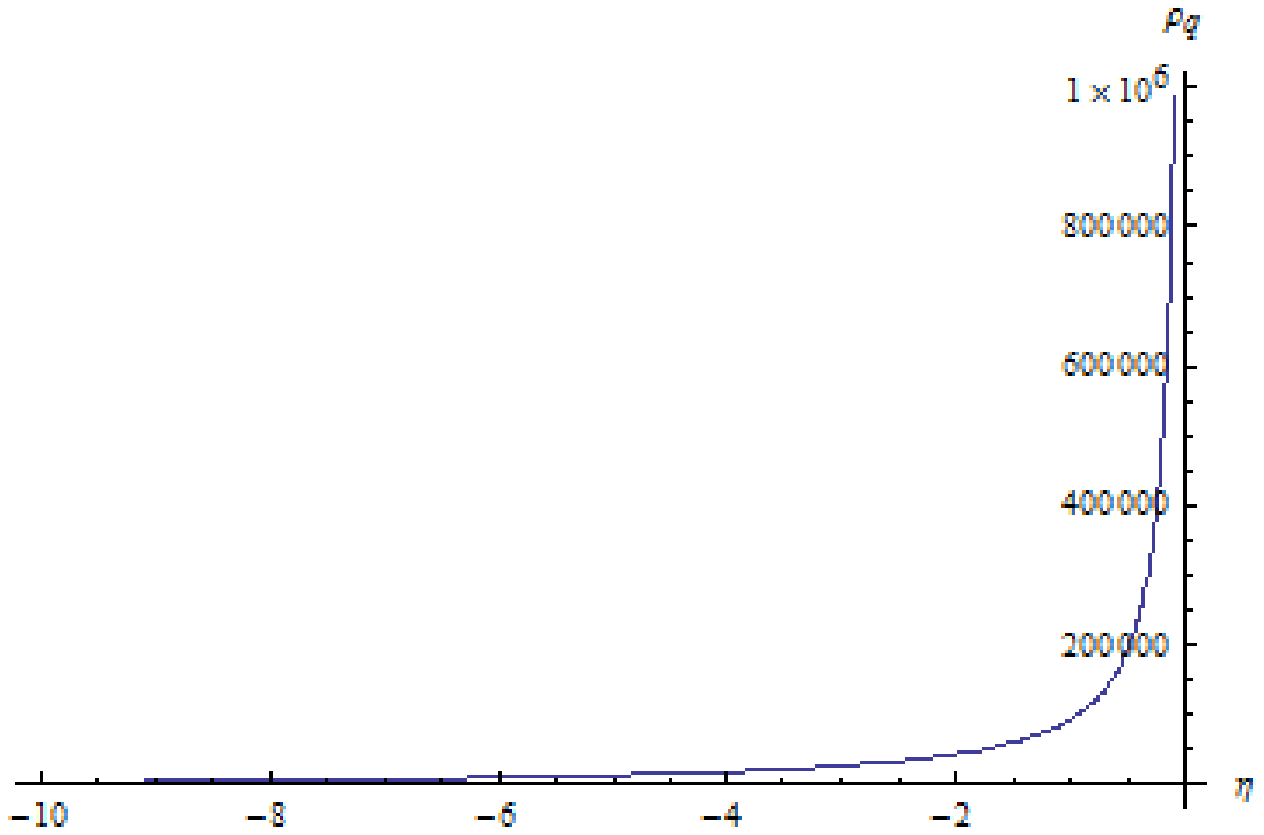}
\end{minipage} \hfill
\begin{minipage}[t]{0.45\linewidth}
\includegraphics[width=\linewidth]{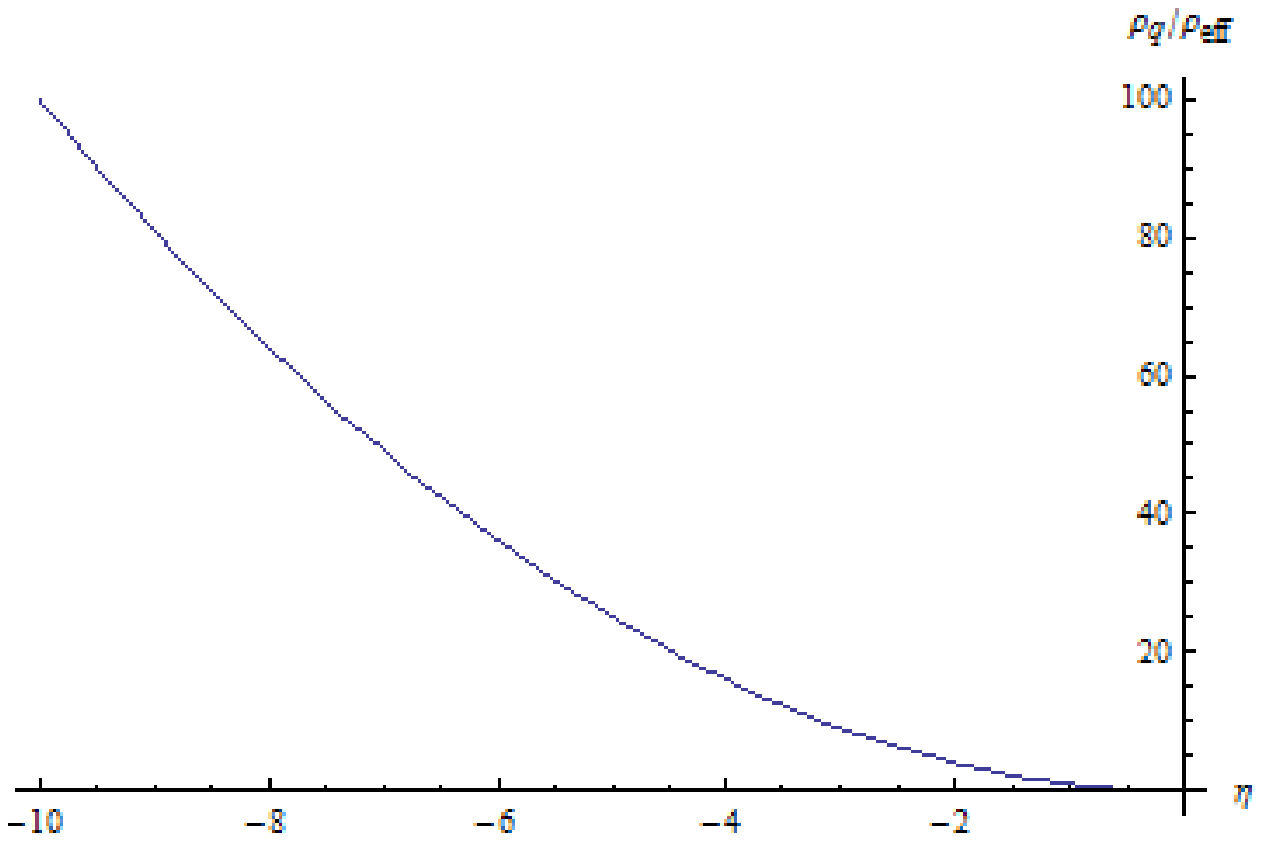}
\end{minipage} \hfill
\begin{minipage}[t]{0.45\linewidth}
\includegraphics[width=\linewidth]{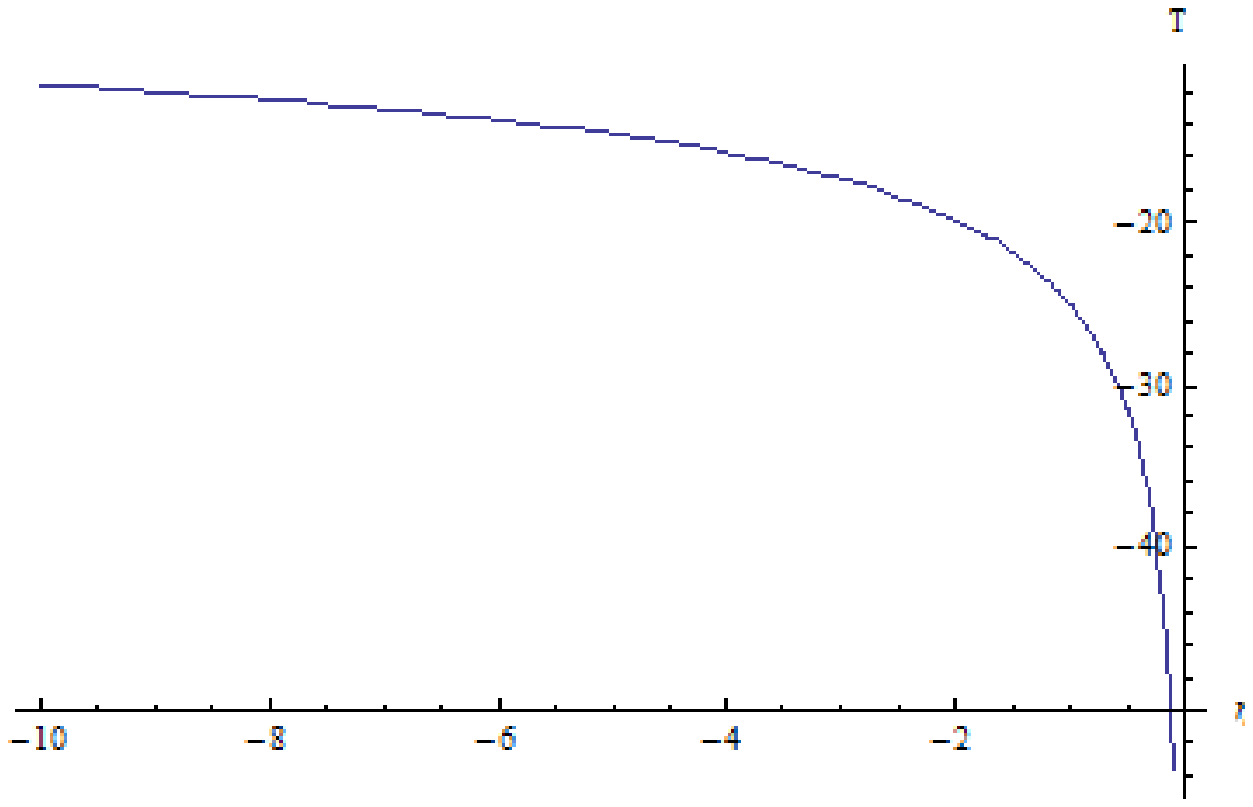}
\end{minipage} \hfill
\caption{{\protect\footnotesize 
The upper left graph shows the evolution of the effective classical energy density of the background and the upper right graph shows the evolution of the quantum energy density due to particle production, both initially small for large values of the conformal time $\eta$, and as the singularity time is approached, their diverge. The lower left graph shows the evolution of the quotient of quantum energy density over the effective classical one, which decreases as the time evolves and vanishes as the singularity is approached. The lower right graph corresponds to the evolution the  torsion scalar in phantom phase, which is initially null and decreases as the time evolves, going toward $-\infty$ as the singularity is approached.}}
\label{}
\end{figure}

\end{center}

\end{document}